\documentclass[%
 reprint,
superscriptaddress,
nobibnotes,
 amsmath,
 amssymb,
 aps,
prb,
]{revtex4-1}
\usepackage{graphicx}
\usepackage[T1]{fontenc}
\usepackage{units}
\usepackage{dcolumn}
\usepackage{bm}

\usepackage{multibib}

\begin{document}

\title{Rectification by hydrodynamic flow in an encapsulated graphene Tesla valve}

\author{Johannes Geurs}
\affiliation{Max Planck Institute for Solid State Research, 70569 Stuttgart, Germany}
\author{Youngwook Kim}
\affiliation{Max Planck Institute for Solid State Research, 70569 Stuttgart, Germany}
\affiliation{Department of Emerging Materials Science, DGIST, 42988 Daegu, Korea}
\author{Kenji Watanabe}
\affiliation{Research Center for Functional Materials, National Institute for Materials Science,\\ 1-1 Namiki, Tsukuba 305-0044, Japan}
\author{Takashi Taniguchi}
\affiliation{International Center for Materials Nanoarchitectonics, National Institute for Materials Science,\\ 1-1 Namiki, Tsukuba 305-0044, Japan}
\author{Pilkyung Moon}
\affiliation{Arts and Sciences, NYU Shanghai, Shanghai 200122, China and NYU-ECNU Institute of Physics at NYU Shanghai, Shanghai 200062, China}
\affiliation{State Key Laboratory of Precision Spectroscopy, East China Normal University, Shanghai 200062, China}
\author{Jurgen H. Smet}
\email{j.smet@fkf.mpg.de}
\affiliation{Max Planck Institute for Solid State Research, 70569 Stuttgart, Germany}

\date{\today}

\begin{abstract}
Systems in which interparticle interactions prevail can be described by hydrodynamics. This regime is typically difficult to access in the solid state for electrons. However, the high purity of encapsulated graphene combined with its advantageous phonon properties make it possible, and hydrodynamic corrections to the conductivity of graphene have been observed. Examples include electron whirlpools, enhanced flow through constrictions as well as a Poiseuille flow profile. An electronic device relying specifically on viscous behaviour and acting as a viscometer has however been lacking. Here, we implement the analogue of the Tesla valve. It exhibits nonreciprocal transport and can be regarded as an electronic viscous diode. Rectification occurs at carrier densities and temperatures consistent with the hydrodynamic regime, and disappears both in the ballistic and diffusive transport regimes. In a device in which the electrons are exposed to a Moir{\'e} superlattice, the Lifshitz transition when crossing the Van Hove singularity is observed in the rectifying behaviour.
\end{abstract}

\maketitle

Hydrodynamics is the study of the collective motion within fluids. By identifying conserved quantities such as mass, momentum and energy, the dynamics of a large variety of systems can be predicted, regardless of the microscopic origin of the interactions within the fluid \cite{Lifshitz1987}. The treatment of interacting electrons in a solid in this manner is referred to as electron hydrodynamics. It serves as a fertile ground for theorists, and electron hydrodynamics has been connected to open research questions such as quantum criticality \cite{Damle1997, Sachdev2008} and the strange metal phase of cuprate superconductors \cite{Zaanen2019}.

In a hydrodynamic system, viscosity is an emergent phenomenon. It determines how transversal momentum diffuses through the fluid and greatly influences the flow.
A prerequisite for viscous flow as a hallmark of hydrodynamics is the global conservation of momentum in the electron fluid \cite{Levitov2015}. This regime tends to be difficult to access  in the solid state context, where electrons readily scatter on impurities or phonons and exchange momentum with the environment. Both of these lead to diffusive transport instead.
Ballistic transport represents the other extreme where all scattering mechanisms are removed. Momentum is then conserved for individual electrons but no collective response can occur.
For electron-electron collisions, momentum is exchanged among particles and total momentum is preserved within the electron liquid \cite{Gurzhi1968a}. Hence, electron viscosity manifests if electron-electron collisions are the dominant source of scattering, i.e. when $l_{ee}\ll W \ll l_{imp}, l_{ph}$. Here, $W$ refers to the sample dimension and $l$ to the electron-electron (ee), impurity (imp) or phonon (ph) scattering lengths. Early efforts in this subfield concentrated on hot electrons in a AlGaAs heterostructure, as these couple minimally to the lattice \cite{Molenkamp1994}. The authors argued that their observations are a manifestation of the Gurzhi effect \cite{Gurzhi1968, Gurzhi1968a}, an anomalous drop of resistance with increasing temperature attributed to Poiseuille flow of an electron liquid.

The advent of graphene has revived the interest in the study of electron hydrodynamics in the solid state as the regime is readibly accessible \cite{Lucas2018, Polini2020}: its large stiffness suppresses electron-phonon interactions and the band structure prohibits Umklapp processes, which exchange momentum between the carriers and the lattice. Moreover, encapsulation in boron nitride (hBN) drastically reduces disorder from impurities as well as strain fluctuations. All of these characteristics open a ``viscous window'' \cite{Ho2018} at intermediate temperature away from the charge neutrality point (CNP) where electron-electron scattering is dominant and electron hydrodynamics can be observed in graphene. There is also an upper limit in carrier density where hydrodynamic effects can be observed. since with larger momentum of the charge carriers at the Fermi energy more and more acoustic phonons become accessible.

The observation of negative nonlocal resistance \cite{Bandurin2016}, attributed to whirlpools forming near a current injection site \cite{Levitov2015}, was the first reported indication of hydrodynamics in graphene. Possible geometric \cite{Pellegrino2016} or ballistic \cite{Shytov2018} contributions make the attribution less certain. A contribution to the conductivity of a point contact in graphene \cite{KrishnaKumar2017} was also attributed to electron viscosity. Direct probes of the electron drift velocity \cite{Sulpizio2019, Ku2020} have also confirmed the Poiseuille flow profile of electrons in graphene. However, a device that relies on the viscous behaviour of electrons, and thereby functions as viscometer, has been missing \cite{Lucas2018}.

\begin{figure}
\begin{center}
\includegraphics{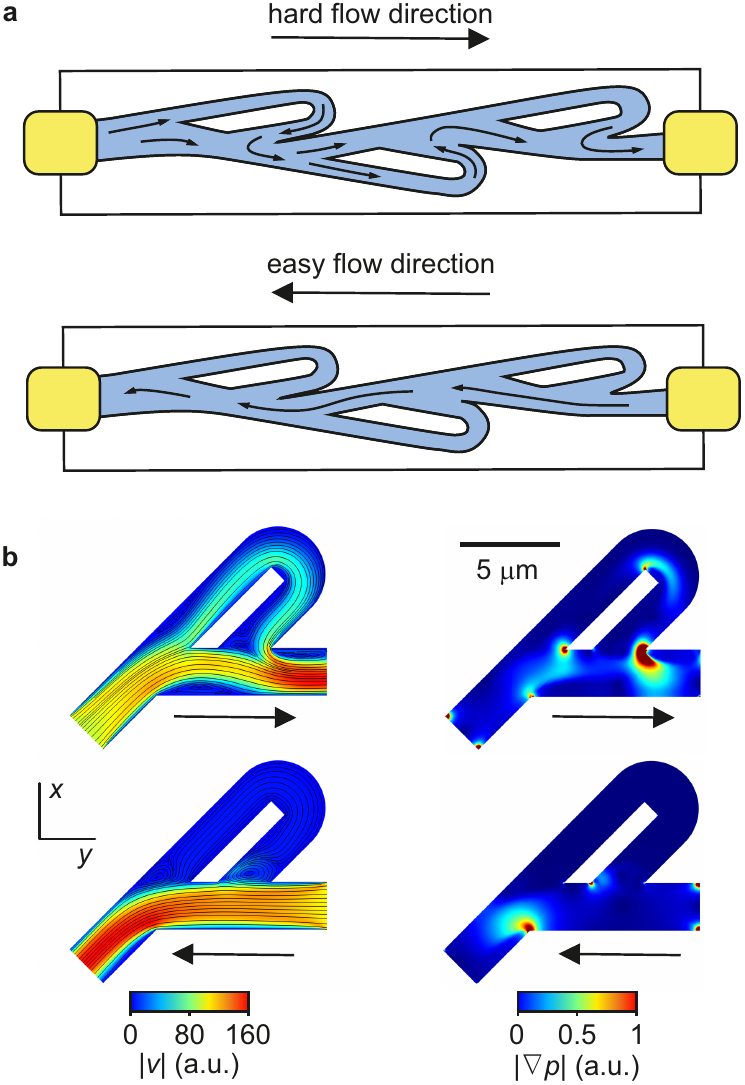}
\caption{\textbf{Mechanical Tesla valve geometry and simulation.} {\bf a}, Original design for the mechanical Tesla valve (modified from Ref. \onlinecite{Tesla1920}). The flow patterns are shown for the hard and easy directions (indicated by the arrows). {\bf b}, Fluid simulations demonstrate the Tesla valve mechanism. The fluid velocity (left column) and pressure change (right column) are shown for both fluid directions (arrows). Significant dissipation occurs where some of the streamlines (black lines) are redirected. The pressure change is calculated as $\left|\nabla p\right|$.}
\label{fig:1}
\end{center}
\end{figure}

Here, we investigate nonreciprocal transport due to viscous electron flow in a  geometry inspired by a mechanical device designed by Tesla \cite{Tesla1920}, commonly referred to as the Tesla valve. The basic component of this valve is a main conduit and a loop departing from and recombining with it at a small angle as illustrated in Fig. \ref{fig:1}a. The schematic shows a device with three such loops in series. A fluid flowing through this device will encounter very different resistance for opposite flow directions. In one direction (top panel), the flow bifurcates and the two streams recombine further along the conduit at a large angle, causing a loss of energy. A simulation of the resulting velocity field and pressure drop, obtained with a commercial fluid dynamics programme, is plotted in the top left and right panel of Fig. \ref{fig:1}b. Clearly visible is a compression of the flow lines when the fluid leaving the loop hits the main flow at an angle. The increased velocity gradient causes flow resistance and dissipation in the viscous fluid, as apparent from the loss of fluid pressure in the panel on the right \cite{Bardell2000}. In the opposite or easy direction, bifurcation is suppressed since the angle of furcation is much larger. Essentially no fluid passes through the loop and the flow resistance is low (bottom panels of Fig. \ref{fig:1}b). Placing multiple loops in series enhances the directional difference in flow resistance and the device turns into a one-way valve without moving parts.

\begin{figure*}
\begin{center}
\includegraphics{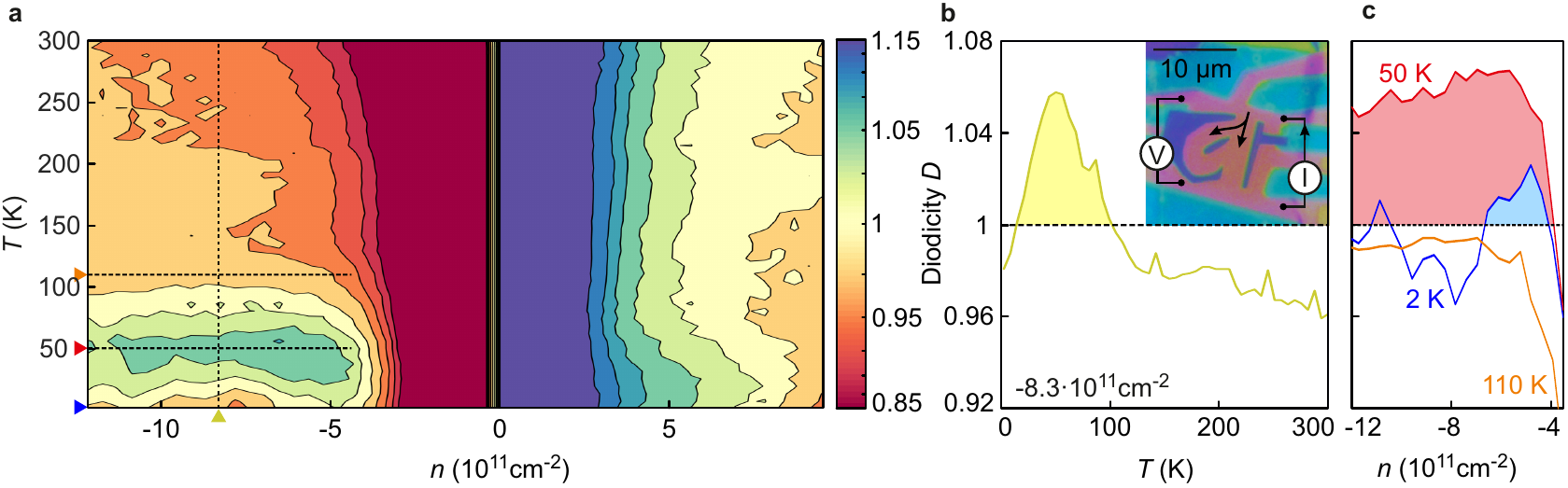}
\caption{\textbf{Rectification in an electronic Tesla valve.} {\bf a}, The ratio of forward and backward resistances (diodicity) as a function of carrier density and temperature for the first device. {\bf b}, Cut at a constant hole density $n$  of $-8.3\cdot 10^{11}$ cm$^{-2}$. {\bf c}, Line traces recorded at constant temperature (2 K, 50 K and 110 K). Shaded areas in panels {\bf b} and {\bf c} highlight regions where carrier viscosity is responsible for the observed change in diodicity. The Inset in panel {\bf b} is an optical micrograph of the first device. The four terminal measurement configuration is shown. Arrows highlight hole flow in the hard direction.}
\label{fig:2}
\end{center}
\end{figure*}

The absence of any moving parts makes the Tesla valve geometry suitable to adaptation in a two-dimensional electron system. State-of-the-art hBN-encapsulated graphene with a low disorder level can meet the requirements for viscous electron transport \cite{Ho2018}. The inset of Fig. \ref{fig:2}b shows the first device, an appropriately patterned graphene flake sandwiched between hBN layers. The van der Waals stack also includes a graphitic backgate. Details of the van der Waals stacks and their fabrication are deferred to the Methods section. Only devices with a single loop are considered. This is sufficient to capture nonreciprocal transport. Note that in a lumped-element model, the rectification ratio does not depend on the number of loops. The degree of rectification is the key figure of merit. It is quantified by the diodicity \cite{Bardell2000}, which we define here as the ratio of the forward and backward resistance, obtained in a four-terminal configuration with constant dc current in either direction ($I_f=-I_b=I$):

\begin{equation}
D=\frac{R_{f}}{R_{b}}=\frac{(V/I)_{f}}{(V/I)_{b}}.
\label{eq:D}
\end{equation}

\noindent We adopt the convention that a positive forward current ($I_f>0$) corresponds to positive charge carrier motion in the ``hard'' direction (indicated by the arrows in the inset of Fig. \ref{fig:2}b). In this direction, some charge carriers will proceed along the loop of the Tesla valve and cause energy dissipation upon rejoining the main stream. With this convention, the diodicity $D$ exceeds 1 for carriers with positive charge. A current carried by negatively charged particles would result in $D<1$, instead. Other possible definitions of the diodicity, such as the ratio of differential resistances, are discussed in the Supplementary Information. The conclusions drawn here remain valid irrespective of the chosen diodicity definition.

Fig. \ref{fig:2}a displays a colour rendition of the diodicity $D$ as a function of the sample temperature $T$ and carrier density $n$ measured on the first device. A positive density $n$ signifies a net density of electron-like charge carriers. The largest deviations from $D=1$ are found near the charge neutrality point (CNP, $n=0$). In this regime, the diodicity barely depends on temperature, suggesting this is not the result of viscous charge carrier flow in the bifurcation. Instead, around the CNP the gate voltage is comparable in magnitude with the applied voltage bias across the device \cite{Meric2008}. As a result of the source-drain voltage drop across the device, the local carrier density becomes a function of the position as well as of the sign and magnitude of the applied current. This is reminiscent of the spatially varying carrier density in the channel of a field-effect transistor and causes unintended rectification irrespective of temperature and the transport regime. The size of the effect scales inversely with the applied gate voltage. A model for this effect has been developed in the Supplementary Information and describes the observations well. Away from the CNP this effect produces a gradually varying and uninteresting background to the diodicity.

On this background behaviour of trivial origin, one area with $D\approx 1.05$ stands out. It encompasses intermediate temperatures ($20K<T<100K$) and large hole concentrations ($n<-5\cdot 10^{11}$ cm$^{-2}$). In Fig.~\ref{fig:2}b we plot a cut through the data at a constant hole concentration of $n=-8.3\cdot 10^{11}$ cm$^{-2}$ highlighting this regime. Fig. \ref{fig:2}c depicts line traces recorded at a constant temperature of 2 K, 50 K and 110 K at these large hole densities. A consistent enhancement of $D$ above 1 is present only in the 50 K trace, while such an increase is absent in the others. The values below 1 in panel b and c are caused by the bias-induced and spatially dependent carrier depletion as described before and in the Supplemental Information.
We attribute the observed change in the diodicity to values above unity to viscous hole flow. Its appearance at intermediate temperatures, away from the CNP supports this claim as it matches the window where hydrodynamic effects have been predicted for graphene \cite{Ho2018}. This is the regime in which the Tesla valve functions as described in Fig. \ref{fig:1}. The flow of positive charge carriers is hampered in the ``hard'' direction of the Tesla valve as seen in the increase of the diodicity. This mechanism is only present in a viscous fluid. Transport at high temperature (110 K trace in Fig. \ref{fig:2}c) is purely diffusive and the device acts as an Ohmic resistor. At low temperature (2 K trace in Fig. \ref{fig:2}c), transport in the device is ballistic. Ballistic rectifiers out of GaAs and graphene~\cite{Song1998, Auton2016} as well as ballistic mesoscopic devices exhibiting transistor-like switching behavior \cite{Wang2019a} have been reported. These devices require multiple submicron orifices or quantum point contacts. Their operation either exploits collimation and reflection of a narrow injected electron stream or an imbalance in the number of supported conduction channels in the different leads with increasing energy \cite{Xu2001, DeHaan2004}. The dimensions of the Tesla valve geometry with micron sized leads are incompatible with these phenomena and therefore these ballistic mechanisms can be safely discarded as the origin of the observed diodicity features in the Tesla valve.

The area in parameter space spanned by the density and temperature where we observe viscous rectification in Fig. \ref{fig:2}a also qualitatively matches the one mapped out in previous work addressing viscous electron flow near current injection contacts \cite{Bandurin2016}. We note that viscous rectification is barely, if at all, visible for the electron side ($n>0$) in our device. Previous studies of graphene heterostructures \cite{Ponomarenko2013, Wang2015, Bandurin2016} generally report better transport characteristics on the hole side, but in theory \cite{Ho2018} the viscous response should be symmetric for both carrier types. Across several devices, the presence of viscous effects depends strongly on the sample quality and size. Narrower channels are influenced more by edge roughness, which destroys momentum conservation in the electron fluid.

\begin{figure*}
\begin{center}
\includegraphics{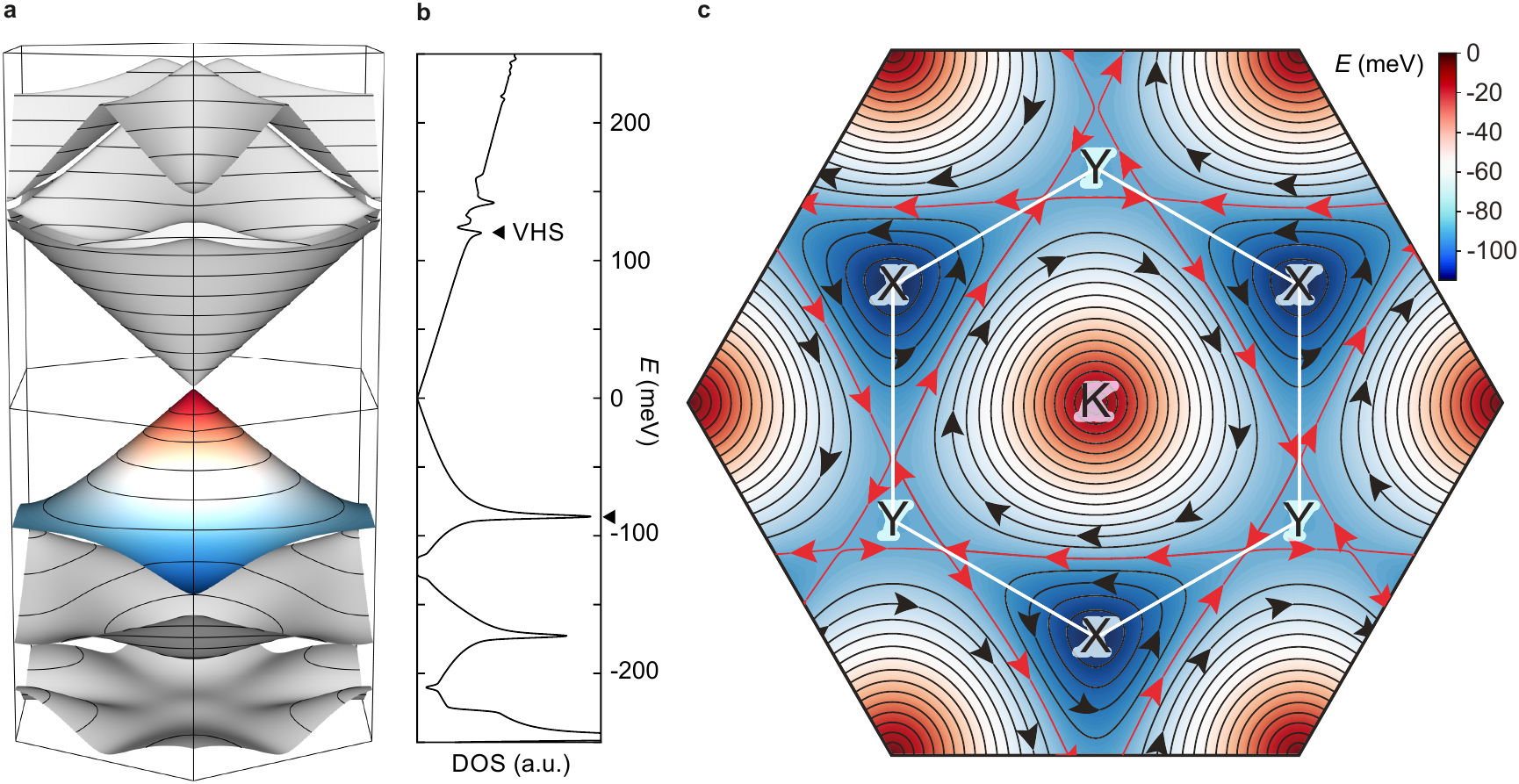}
\caption{\textbf{Band structure of monolayer graphene with a Moir\'{e} superlattice potential.} {\bf a}, The energy dispersion in the reduced Brillouin zone of the miniband centered around zero energy closely resembles the band structure of freestanding graphene, but on a greatly reduced energy scale. {\bf b}, Calculated density of states. Note the Van Hove singularity peaks for the hole side ($E<0$) and for the electron side ($E>0$). {\bf c}, Contour plot for the colored, negative energy portion of the lowest miniband  in panel a. The arrows indicate the chirality of the Fermi contours. Close to zero energy, the hole-like carriers circle around $K$. At lower energy, a Lifshitz transition occurs and the carriers become electron-like and encircle the $X$ symmetry point of the reduced Brillouin zone.}
\label{fig:3}
\end{center}
\end{figure*}

In a second experiment, a monolayer of graphene with a Moir{\'e} superlattice \cite{Ponomarenko2013} induced by the neighbouring hBN layer was patterned into a Tesla valve. Zone folding by the superlattice period gives rise to a rearrangement of the electronic band structure into minibands as illustrated in Fig. \ref{fig:3}. Panel a shows the miniband centered around $K$. At zero energy, it resembles the $\pi$ and $\pi^{*}$ band structure of graphene except for a dramatic downscaling of the energy. It also includes a Van Hove singularity (VHS) for both positive and negative energy, visible as a peak in the calculated density of states (Fig. \ref{fig:3}b). This peak is very pronounced for negative energy, i.e. hole occupation of the valence portion of the miniband centered around zero energy, whereas it is weak for electron occupation because here the flat area of the saddle point near $+0.13{\rm eV}$ consumes a small area of k-space only. In addition, the second conduction band already overlaps.  The longitudinal resistance is anticipated to peak not only when the chemical potential crosses the Dirac point at zero energy, but also when the miniband is either fully depleted or occupied. Fig.~S3b 
in the Supplementary Information plots the longitudinal resistance as a function of the density and indeed exhibits secondary charge neutrality peaks at high electron and hole concentration signalling full occupation and depletion of the miniband. They serve as unequivocal hallmarks for the presence of the Moir\'{e} superlattice potential and their position unveils the twist angle between graphene and hBN, which is for this sample less than $1^\circ$.

The Van Hove singularity is also associated with a Lifshitz transition. Fig. \ref{fig:3}c shows a contour plot for negative energies of the first miniband around $K$ for a graphene layer with a Moir\'{e} superlattice. At low hole densities (areas in red), the Fermi contours encircle the $K$ symmetry points of the reduced Brillouin zone and charge carriers indeed behave as holes. However, once the chemical potential drops below the Van Hove energy, the Fermi contours encompass the $X$ symmetry point. The positive dispersion of the electronic miniband at energies away from the $X$ point implies that charge carriers no longer behave like holes but rather like electrons. The topological transition of the Fermi surface turns holes to electrons when the ``hole'' doping is increased, to which the Tesla valve should be sensitive.

\begin{figure}
\begin{center}
\includegraphics[width=0.45\textwidth]{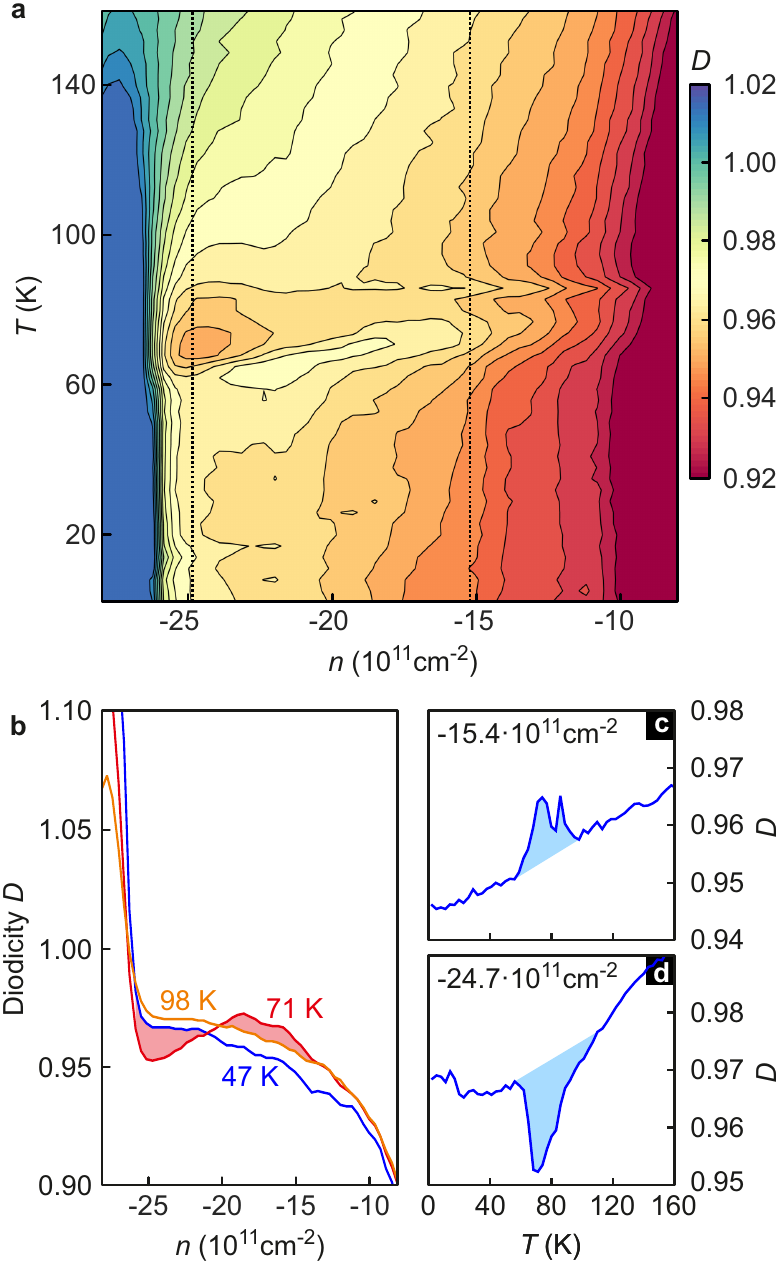}
\caption{\textbf{Diodicity in a Tesla valve with a Moir{\'e} superlattice potential.} {\bf a}, $D(n,T)$ for large hole concentrations. As the primary and secondary charge neutrality points are approached (moving to the right and left, respectively) the diodicity drops  well below 1 (red) or exceeds 1 (blue) due to the source-drain bias-induced spatial dependence of the density. Of interest is not this overall background behaviour, but the behaviour around 70 K, where viscous hole-like carriers will increase the diodicity (area at $n<-12\cdot 10^{11}cm^{-2}$) and viscous electron-like carriers will cause a decrease (area around $n\approx -24\cdot 10^{11}cm^{-2}$) as the chemical potential crosses the Van Hove singularity. {\bf b}, Line traces recorded at constant temperatures. The excursions of the diodicity recorded at 71 K away from the background, represented by the traces taken at 47 K or 98 K, are attributed to electron viscosity and shaded in red. {\bf c}, {\bf d}, Line traces at constant density, as marked with dotted lines in panel {\bf a}. Blue areas show the temperature range in which viscous charge carriers cause rectification. The sign of the viscous contribution depends on the carrier species.}
\label{fig:4}
\end{center}
\end{figure}

Fig. \ref{fig:4}a displays the $D(n,T)$ map of this device for hole densities between $n=-8\cdot 10^{11}$ cm$^{-2}$ and $n=-28\cdot 10^{11}$ cm$^{-2}$. As in the first device, the observed behaviour for $D$ is dominated primarily by significant deviations from unity due to source-drain bias-induced carrier density variations. The diodicity reaches an inflection point as we move away from the primary CNP where $D\ll 1$ (red) towards the secondary CNP at $n=-28\cdot 10^{11}$ cm$^{-2}$ (from right to left in Fig. \ref{fig:4}a) where $D\gg 1$ (blue). This creates a smoothly changing background. However, at temperatures between 50 K and 100 K there is an anomalous increase followed by an anomalous decrease of $D$. The line trace extracted at a constant temperature of 71 K (Fig. \ref{fig:4}b) illustrates that $D$ first exceeds and then dives below the background (traces at 47 K and 98 K are representative for the background behaviour). These areas of larger and smaller diodicity are also visible in the two traces recorded at constant carrier densities in panels c and d.
We assert that this more complex behaviour of the diodicity is again a manifestation of viscous carrier flow in conjunction with a change in the Fermi surface topology as the chemical potential crosses the Van Hove singularity. At low hole densities, the viscous flow of the hole-like carriers is hindered in the ``hard direction'', increasing the diodicity in the viscous regime between 50 K and 100 K. However, once the chemical potential traverses the Van Hove singularity, the carriers are electron-like. Accordingly, the particles move in the opposite direction and the diodicity drops below the background.

On this sample with the Moir\'{e} superlattice potential, measurements were also performed prior to etching the Tesla valve geometry up to a density of about $18 \cdot 10^{11}$cm$^{-2}$ where the valve geometry displays viscous rectification. The results on the initial simple square geometry are included in Fig. S4 
of the Supplementary Information (panel a). The device does not show signs of viscous rectification. These only appear after patterning the Tesla valve geometry on this very same van der Waals heterostructure. This corroborates that the observed effect is not an intrinsic material property or the result of contact imperfections, but rather follows from the chosen device geometry. In contrast, the trivial source-drain bias-induced rectification near the CNP (Supplementary Information, Fig.~S1 
is inherently present in every measured device, also in the square geometry, as clearly seen by comparing panel a and b in this figure.
For the sake of completeness we note that as a result of the Moir\'{e} potential-induced band structure changes, Umklapp scattering may occur in this second device. Such scattering destroys hydrodynamic effects, yet viscosity-induced rectification is still observed, albeit in a smaller temperature window in this device with Moir\'{e} potential. It suggests that electron-electron scattering continues to prevail in this temperature regime. Umklapp scattering is strongly dependent on temperature \cite{Wallbank2019,Kim2020} and, to the best of our knowledge, the relative scattering rates of Umklapp processes compared to electron-electron scattering have not been studied. This makes it unclear at which temperature Umklapp scattering takes the lead.

In conclusion, we have studied viscosity-induced nonreciprocity in encapsulated monolayer graphene devices. The Tesla valve geometry offers flow direction dependent dissipation and can be regarded as an electronic viscometer via the strength of the rectification. This method also demonstrated viscous behaviour of secondary Dirac electrons in a Moir\'{e} superlattice. The marriage of fluid concepts with the readily accessible hydrodynamic flow regime in graphene with its exceptional tunability and patternability is bound to be a fertile playground for future basic research in the rich field of fluid mechanics.

\section*{Methods}

The van der Waals heterostructures in this work consist of four layers. A graphene monolayer is sandwiched between two hBN layers with thicknesses between 50 nm and 80 nm. Each stack also rests on top of a graphite layer with a typical thickness of a few nm. It serves as the backgate to control the carrier density of the graphene.

The van der Waals stacks were created with the dry pickup method \cite{Castellanos-Gomez2014}. The constituent hBN and graphene flakes were mechanically exfoliated \cite{Novoselov2004} on the 300 nm thermally grown SiO$_2$ layer of Si substrates. Suitable flakes were chosen based on their lateral dimensions, thickness and flatness. The flakes were picked up with a polymer (Elvacite 2552C, Lucite International) drop that was formed on a glass slide. The glass slide was mounted on a micromanipulator. The substrates were heated to a temperature of 110$^\circ$C during pickup to improve adhesion to the stamp. After all flakes were picked up, the entire stack was released onto a Si substrate. Remaining polymer residues were removed during a subsequent annealing step in forming gas for 10 minutes at 500$^\circ$C. For clean layers this annealing step frequently causes a rotation of the 2D layers towards a lower energy configuration \cite{Wang2015}. This gives rise to a Moir{\'e} superlattice potential. The second device discussed in the text is such an example.

The van der Waals stacks were patterned into the single loop Tesla valve geometry with the help of electron beam lithography. Exposed areas were treated with plasma etching. To etch hBN a SF$_6$/Ar plasma was used, whereas graphene was removed with an O$_2$ plasma. Electrical contacts to the graphene were also fabricated by electron beam lithography in conjunction with thermal evaporation of Cr (10 nm) and Au (50 nm). The active area of the completed device was treated by AFM ironing, a technique that releases strain and improves the overall transport characteristics \cite{Kim2019a}.

The transport measurements were performed in a physical property measurement system (QuantumDesign DynaCool PPMS) in which the sample temperature can be controlled between 2 K and 300 K. The carrier density was determined from the periodicity of Shubnikov-de Haas oscillations and the position of integer quantum Hall features.

The transport data reported in the main text were acquired with DC excitation at zero magnetic field. A total of 2550 I-V characteristics were recorded on device 1 and 14100 on device 2. The applied current was 6.3 $\mu$A for device 1, 8.3 $\mu$A for device 2 (square) and 690 nA for device 2 (patterned). The corresponding voltage $V$ in forward and backward current directions was then used to calculate the diodicity according to Eq. \ref{eq:D}.

The simulation of the hydrodynamic flow in a mechanical Tesla valve was carried out with the COMSOL simulation package. The following parameters were chosen for the simulation: $\rho$=1 kg/m$^3$, viscosity $\mu$=1 $\mu$Pa $\cdot$ s and inlet velocity 100 m/s.

\section*{Acknowledgements}

The authors thank Hans Boschker, Shahal Ilani, Hyeonjeong Kim, Klaus von Klitzing and Tim Rowett for insightful discussions.
J.H.S.~acknowledges financial support from the graphene flagship core 3 program.
The work at DGIST was supported by the Basic Science Research Program NRF-2020R1C1C1006914 through the National Research Foundation of Korea (NRF) and also by the DGIST R\&D program (20-CoE-NT-01), funded by the Korean Ministry of Science and ICT.
P.M. acknowledges the support from Science and Technology Commission of Shanghai Municipality grant no. 19ZR1436400, and NYU-ECNU Institute of Physics at NYU Shanghai. This research was carried out on the High Performance Computing resources at NYU Shanghai.
The growth of hexagonal boron nitride crystals was sponsored by the Elemental Strategy Initiative conducted by the MEXT, Japan, Grant Number JPMXP0112101001, JSPS KAKENHI Grant Number JP20H00354
and the CREST(JPMJCR15F3), JST.

\section*{Author contributions}

J.G. designed and carried out the experiments. J.G. and Y.K. fabricated the graphene devices. K.W. and T.T. synthesized the hBN crystals. P.M. provided the theory. J.G., Y.K. and J.H.S. wrote the manuscript.



\pagebreak

\onecolumngrid
\begin{center}
  \textbf{\large Supplementary Information for ``Rectification by hydrodynamic flow in an encapsulated graphene Tesla valve''}\\[.2cm]
  Johannes Geurs$^{1}$, Youngwook Kim$^{1,2}$, Takashi Taniguchi$^{3}$, Kenji Watanabe$^{4}$, Pilkyung Moon$^{5,6}$, Jurgen H. Smet$^{1*}$\\[.1cm]
  {\itshape ${}^1$Max Planck Institute for Solid State Research, 70569 Stuttgart, Germany\\
  ${}^2$Department of Emerging Materials Science, DGIST, 42988 Daegu, Korea\\
  ${}^3$Research Center for Functional Materials, National Institute for Materials Science,\\ 1-1 Namiki, Tsukuba 305-0044, Japan\\
  ${}^4$International Center for Materials Nanoarchitectonics, National Institute for Materials Science,\\ 1-1 Namiki, Tsukuba 305-0044, Japan\\
  ${}^5$Arts and Sciences, NYU Shanghai, Shanghai 200122, China and\\ NYU-ECNU Institute of Physics at NYU Shanghai, Shanghai 200062, China\\
  ${}^6$State Key Laboratory of Precision Spectroscopy, East China Normal University, Shanghai 200062, China\\}
  ${}^*$Electronic address: j.smet@fkf.mpg.de\\
(Dated: \today)\\[1cm]
\end{center}

\setcounter{equation}{0}
\setcounter{figure}{0}
\renewcommand{\theequation}{S\arabic{equation}}
\renewcommand{\thefigure}{S\arabic{figure}}
\renewcommand{\bibnumfmt}[1]{[S#1]}
\renewcommand{\citenumfont}[1]{S#1}

\section*{S1: Transistor models}

\begin{figure}
	\begin{center}
		\includegraphics{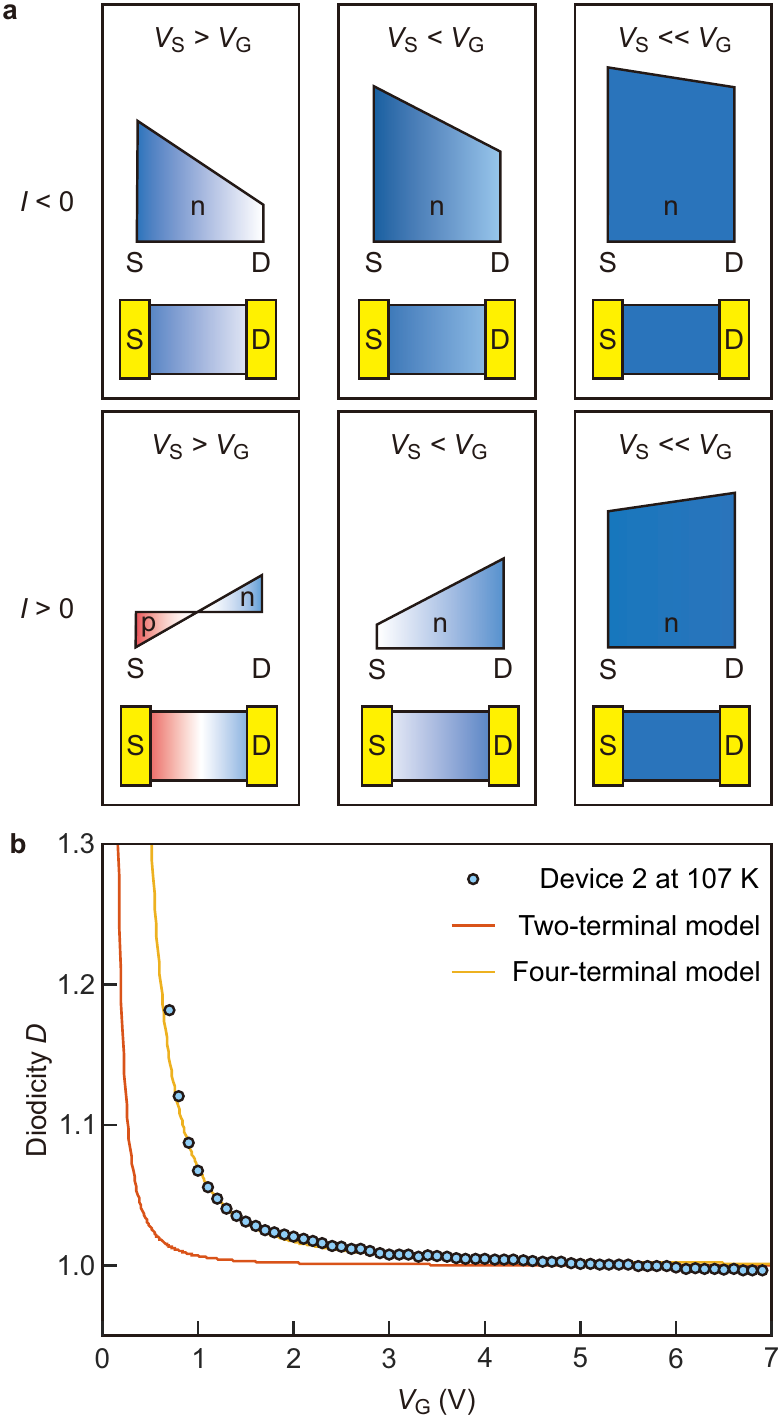}
		\caption{\textbf{Bias-induced spatial dependence of the carrier density.} {\bf a}, Top: Schematic of carrier concentration with varying source-drain bias and gate voltage when negative current is applied. Bottom: The same as top panel but with positive current. When the source-drain voltage $V_S$ becomes comparable to the gate voltage $V_G$, the carrier concentration can vary strongly along the device and even change polarity. {\bf b}, Comparison of the two- and four-terminal models for the bias induced diodicity (Eqs. \ref{eq:2T} and \ref{eq:4T}). Also shown are experimental data points recorded on Device 2 at 107 K, outside of the viscous regime.}
		\label{fig:S1}
	\end{center}
\end{figure}

\noindent The mechanism for a bias-induced spatial dependence of the carrier density is illustrated schematically in Fig. \ref{fig:S1}a. The carrier concentration in a device (green area) is determined by the gate voltage $V_G$. When the drain-source voltage $V_S$ becomes comparable in magnitude (left panels), the carrier concentration and conductivity will vary spatially. This is identical to what occurs in MOSFET devices \cite{Sze2006}. As opposed to conventional MOSFETs, a large enough bias can even induce carriers of the opposite sign in the channel \cite{Meric2008} due to the lack of a band gap in graphene.
To model the effect, we consider a two-terminal transistor first. A current $I$ flows in the channel with width $W$ and length $L$ between the source contact $S$ at potential $V_S$ and the drain contact $D$ which is grounded ($V_D=0$). A capacitance $C$ describes the coupling between the channel and the gate at potential $V_G$. Current continuity in the channel yields the standard transistor equation \cite{Sze2006}

\begin{equation}
\frac{IL}{Ce\mu W}=V_G\cdot V_S -\frac{V_S^2}{2},
\label{eq:2T}
\end{equation}

\noindent where $-e$ is the electron charge and $\mu$ is the channel mobility. The red curve in Fig. \ref{fig:S1}b shows the diodicity (Eq. (1) in the main text) 
 for this model. In general, the diodicity does not depend strongly on temperature and scales inversely with $V_G$ for transistor-like models.

This bias-induced carrier change is stronger in a four-terminal setup, as there is an additional potential drop between the voltage probes and the drain. The internal contacts between source and drain, i.e.~the voltage probes, are referred to as contacts 1 and 2. The potential difference between contact 2 and the drain will affect the carrier density between contacts 1 and 2, making the I(V) characteristic more nonlinear. This can be modelled by applying Eq. \ref{eq:2T} to both portions of the transistor channel:
\begin{align}
	I_{12} &=Ce\mu\left(\frac{W}{L}\right)_{12} \left(V_G(V_1-V_2)-\left(\frac{V_1^2}{2}-\frac{V_2^2}{2}\right)\right) \\
	I_{2D} &=Ce\mu \left(\frac{W}{L}\right)_{2D} \left(V_G(V_2-V_D)-\left(\frac{V_2^2}{2}-\frac{V_D^2}{2}\right)\right).
\end{align}
\noindent With $I=I_{12}=I_{2D}$ and $V_D=0$, these expressions can be rearranged to
\begin{equation}
V_1-V_2=\Delta V(I)=\sqrt{V_G^2-K\frac{2I}{Ce\mu}\cdot\left(\frac{L}{W}\right) _{12}}-\sqrt{V_G^2-(K+1)\frac{2I}{Ce\mu}\cdot\left(\frac{L}{W}\right) _{12}}.
\label{eq:4T}
\end{equation}
\noindent Here the shape factor $K=\frac{(W/L)_{12}}{(W/L)_{2d}}$ is a geometric constant, the ratio between the number of squares between contact 2 and the drain, and the number of squares between contacts 1 and 2. Equation \ref{eq:4T} has been used to model the diodicity of a four-terminal transistor, shown as the yellow line in Fig.~\ref{fig:S1}b, with $K=4.5$ chosen as the only fit parameter to the experimental data at 107 K. Note that Eq. \ref{eq:4T} reduces to Eq. \ref{eq:2T} in the case where $K=0$.

This bias-induced carrier change describes the observed diodicity (blue dots in Fig. \ref{fig:S1}b) well. It creates a large and temperature-independent signal that obscures signatures of viscous behaviour particularly at low carrier densities. However, even at higher carrier density it remains present and accounts for an overall gradual change of the diodicity that does not strongly depend on temperature.

\newpage
\section*{S2: Alternative definitions of diodicity}

\begin{figure}
\begin{center}
	\includegraphics{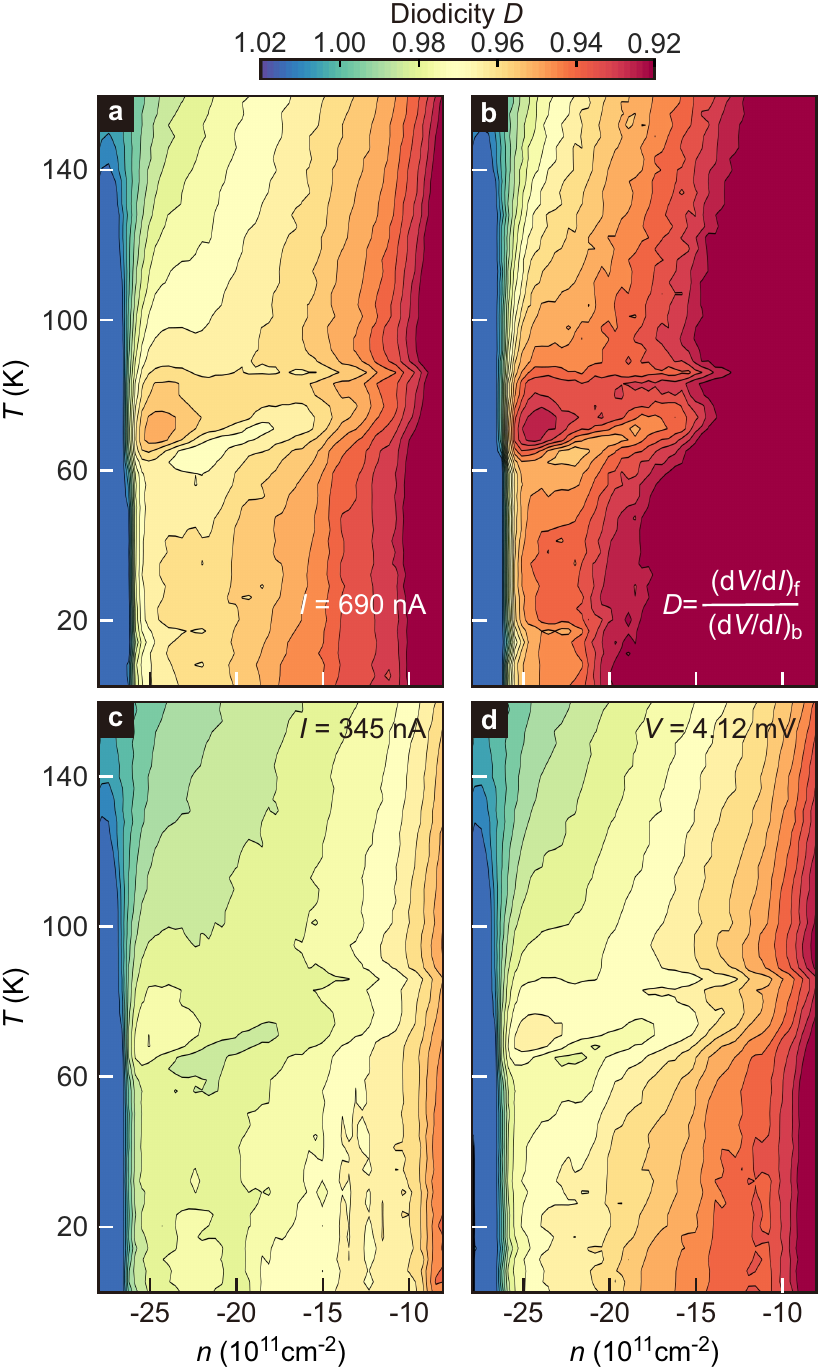}
	\caption{\textbf{Alternative ways to quantify nonreciprocal transport.} {\bf a}, Diodicity calculated according to Eq. 1 in the main text 
		 for $I_f=-I_b=$ 690 nA. {\bf b}, Diodicity based on differential resistances. Data were calculated for $I_f=-I_b=$ 650 nA and an ac current excitation of 90 nA. {\bf c}, The same as panel {\bf a} but for $I_f=-I_b=$ 345 nA.  {\bf d}, Diodicity calculated for a constant voltage $V_f=-V_b=$ 4.12 mV, instead of constant current. All panels share the same colour scale. Note that the qualitative behaviour of $D(n,T)$ does not change.}
	\label{fig:S2}
\end{center}
\end{figure}

Multiple definitions of the figure of merit for nonreciprocity, the diodicity, are possible. The choice is ultimately arbitrary and does not alter any of the conclusions in this manuscript. Some alternatives to Eq. (1) in the main text 
 are illustrated in Fig. \ref{fig:S2} for device 2. At the left panels, the diodicity according to Eq. (1) in the main text 
  is shown for two different current values: $I_f=-I_b=$ 690 nA and 345 nA. In panel b, a definition based on differential resistance is used instead. In panel d, the diodicity was calculated for a constant voltage $V_f=-V_b=$ 4.12 mV instead of a constant current. The salient features of this device (large diodicity near the primary and secondary CNPs as well as the smaller viscous contribution at intermediate temperature that changes sign) are present in all panels.

\newpage
\section*{S3: Secondary Dirac peaks in device 2}

\begin{figure}
\begin{center}
	\includegraphics{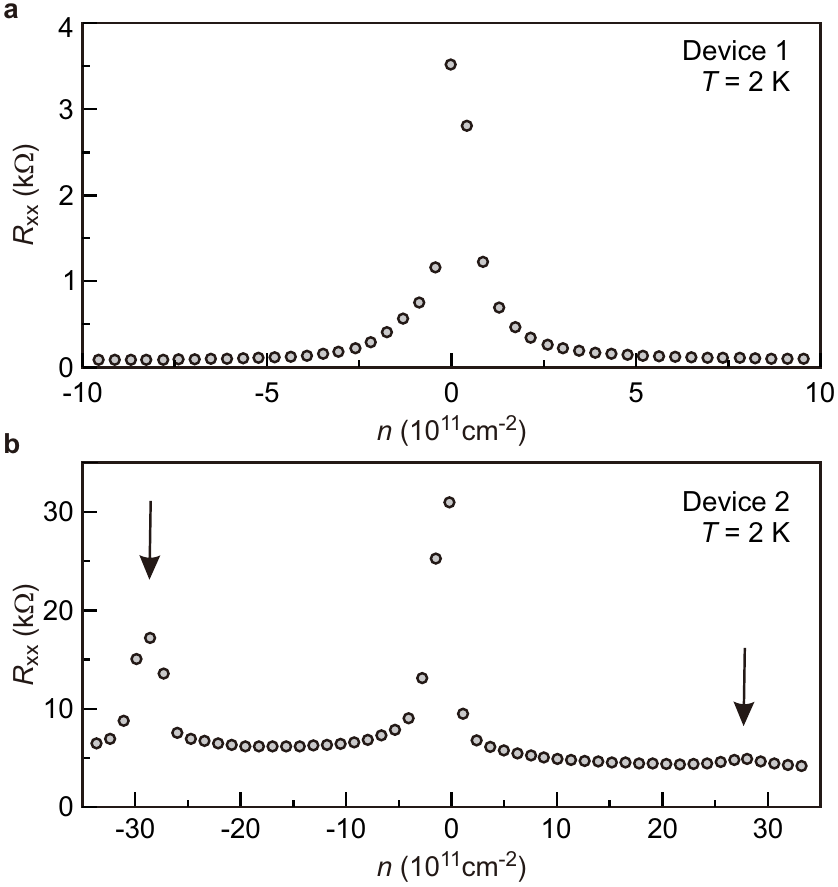}
	\caption{\textbf{Four terminal resistance $R_{xx}$ versus density for both devices.} {\bf a}, Density dependence of longitudinal resistance of Device 1. {\bf b}, The same as {\bf a} but for Device 2. Secondary Dirac peaks of the second device are indicated by the arrows. They are equidistantly separated with respect to the main Dirac peak. The data were recorded at $T$=2 K in the absence of a magnetic field.}
	\label{fig:S3}
\end{center}
\end{figure}

Figure \ref{fig:S3} shows the resistance of both devices discussed in the main text as a function of gate voltage. These traces were recorded after the final patterning step. Two additional charge neutrality peaks in the resistance of the second device due to the Moir\'{e} superlattice potential
are indicated by the arrows. They appear at $n\approx\pm 28\cdot 10^{11}cm^{-2}$, consistent with a graphene/hBN superlattice with a twist angle below $1^\circ$ \cite{Yankowitz2012, Ponomarenko2013}.

\newpage
\section*{S4: Diodicity of a square sample versus Tesla valve sample}

\begin{figure*}
\begin{center}
	\includegraphics{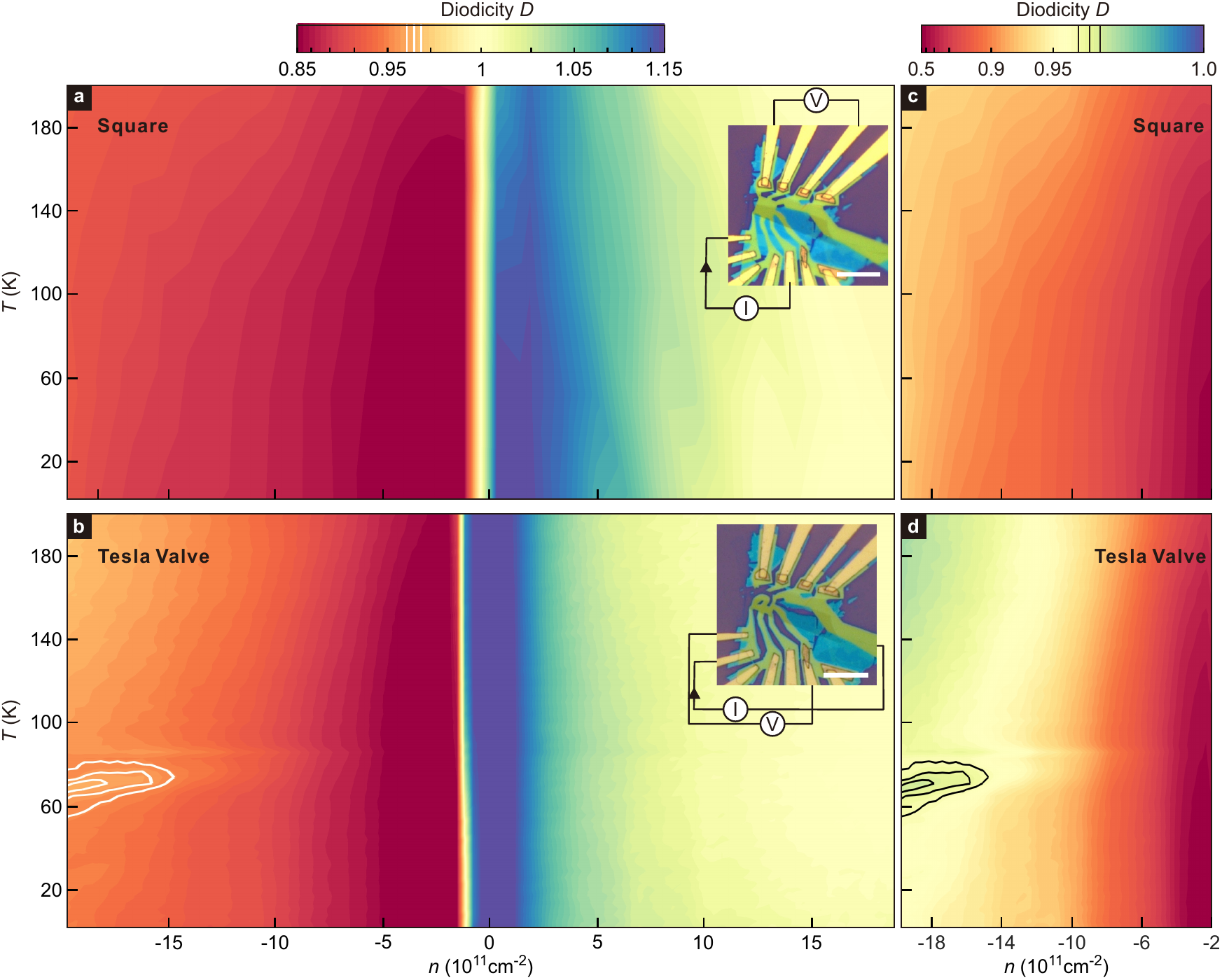}
	\caption{\textbf{Comparison between the diodicity of a device that initially had a square shape and that was subsequently repatterned into the Tesla valve geometry.}  {\bf a}, Diodicity as a function of carrier density and temperature for the square geometry. A large deviation of the diodicity from 1 near the charge neutrality point is the only visible feature. The inset shows device image and measurement setup. The scale bar is 10 $\mu$m. {\bf b}, $D(n,T)$ for the same device, but now etched in the Tesla valve geometry. Colour scale on top of panel {\bf a} is for both {\bf a} and {\bf b}. Measurement set-up and optical image of etched device are displayed in the inset with 10 $\mu$m scale bar. The contour lines emphasize the extra feature around 80 K at large hole density compared with panel {\bf a}. {\bf c},{\bf d}, A zoom in on the area in which the extra feature (black lines) appears. To distinguish the details of rectification in the Tesla valve geometry, a different colour scale shown on top of panel {\bf c}, is introduced for panel {\bf c},{\bf d}.}
	\label{fig:S4}
\end{center}
\end{figure*}
In order to confirm that the rectification observed in the Tesla valve is really geometry-induced and not an intrinsic property of the van der Waals stack or an artefact of -- for instance -- contacts, the device with the Moir\'{e} potential was also measured in a square geometry prior to patterning the Tesla valve geometry itself. The diodicity measurements are summarized in Fig. \ref{fig:S4}a. Only the mostly temperature-independent contribution to the diodicity generated by the source-drain bias-induced spatial dependence of the carrier density appears. This contribution of trivial origin is large near the CNP and the diodicity gradually decreases when moving away from the CNP.

After these measurements, this sample was warmed up and reprocessed by etching the Tesla valve geometry out of the initial square shaped device area. The measurement of the upper panel was then repeated. The $D(n,T)$ diagram has been discussed in Fig.4 
 of the main text. For the sake of comparison, the data are replotted here in the lower panel (Fig.~\ref{fig:S4}b) on the same axis as the upper panel. At large hole density ($n<-10\cdot 10^{11}cm^{-2}$) and intermediate temperature ($50K<T<100K$) an additional feature (white lines) appears that is not present in the upper panel. As discussed in the main text, viscous flow in the Tesla valve, in conjunction with the Lifshitz transition at the VHS, is responsible for this feature.

Panels c and d show the same measurements with a reduced density window, presented at a different colour scale to emphasize the appearance of an additional area of rectification (black lines).

The device, before and after patterning, is visible in the insets, together with the four-terminal measurement setup.

\end{document}